*Research Article*

# On the Analytical Solution of Fractional SIR Epidemic Model

**Ahmad Qazza and Rania Saadeh**

*Department of Mathematics, Faculty of Science, Zarqa University, Zarqa 13110, Jordan*

Correspondence should be addressed to Rania Saadeh; rsaadeh@zu.edu.jo

This article presents the solution of the fractional SIR epidemic model using the Laplace residual power series method. We introduce the fractional SIR model in the sense of Caputo's derivative; it is presented by three fractional differential equations, in which the third one depends on the first coupled equations. The Laplace residual power series method (LRPSM) is implemented in this research to solve the proposed model, in which we present the solution in a form of convergent series expansion that converges rapidly to the exact one. We analyze the results and compare the obtained approximate solutions to those obtained from other methods. Figures and tables are illustrated to show the efficiency of the LRPSM in handling the proposed SIR model.

## 1. Introduction

Epidemiology is a discipline of biology that examines the prevalence and causes health-related problems in particular groups or communities. It may be used to manage community health concerns. Frequency, distribution, and the factors that cause diseases are the elements of epidemiology. By establishing priorities among these services, the Department of Epidemiology seeks to offer the data required for the planning, execution, and assessment of services aimed at disease prevention, control, and treatment. Epidemiology is used to investigate the historical development and decline of illnesses in the population, community diagnosis, planning and evaluation, assessment of an individual's risks and chances, identification of syndromes, and completion of the disease's natural history [1–5].

The study of fractional calculus depends on computing integrals and derivatives of noninteger orders. The importance of these studies has appeared in the various applications in physics and engineering, in which these derivatives can describe them more realistically in the fields of science. A more accurate explanation of the challenges in the actual world can be provided by the fractional application models [6–11].

Through the process of mathematical modeling, one may examine, anticipate, and offer insight into issues that arise in the actual world. It is advantageous since conventional theoretical approaches are inadequate for the study of technological, ecological, economic, and other systems examined by modern research. Fractional differential equations and systems have been used to simulate a variety of real-world issues [12–18].

The SIR model is one of the simplest fractional models, and many models are derived depending on its form. The model consists of three components:

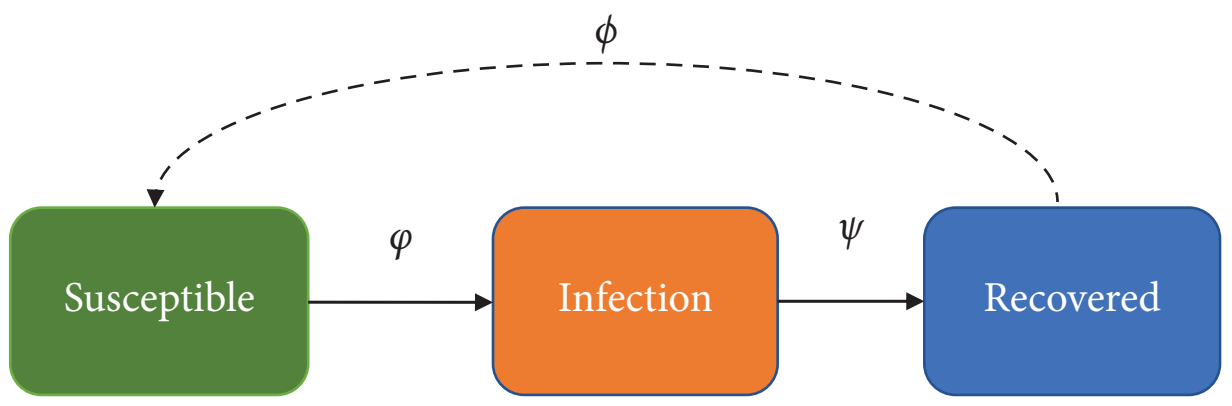

$\varphi$: The number of susceptible individuals. When a susceptible and an infectious individual come into "infectious contact," the susceptible individual contracts the disease and transitions to the infectious side.

$\psi$: The number of infectious individuals. These are individuals who have been infected and are capable of infecting susceptible individuals.

$\phi$: The number of removed (and immune) or deceased individuals.

The study of the SIR model was first introduced in 2009 by Ahmet and Cherruault [19], then in 2011 they present new research on analytical solutions of some related models. After that, many authors have investigated the susceptible-infected recovered models of integer fractional orders [20–23].

The integer-order SIR epidemic model is given by

$$\begin{aligned}\varphi'(\tau) &= -q_1\varphi(\tau)\psi(\tau) + q_3\psi(\tau),\\ \psi'(\tau) &= q_1\varphi(\tau)\psi(\tau) - q_3\psi(\tau) - q_2\psi(\tau),\\ \phi'(\tau) &= q_2\psi(\tau),\end{aligned} \tag{1}$$

under the initial conditions:

$$\begin{aligned}\varphi(0) &= \alpha,\\ \psi(0) &= \beta,\\ \phi(0) &= \lambda,\end{aligned} \tag{2}$$

where $\varphi(\tau)$ denotes the susceptible individuals, $\psi(\tau)$ denotes the infected individuals, $\phi(\tau)$ denotes the removed individuals, and $\tau$ denotes the time. $N$ is the total number of the studied population such that $\varphi(\tau) + \psi(\tau) + \phi(\tau) = N$. The rate of changes between the previous and the new values are $q_1, q_2$ and $q_3$, for more details see [24, 25]. There are some constraints on the model concerning the population number N, which should be large enough, the parameters of the system should be fit, and finally, the healing does not provide immunity.

This research presents the LRPSM, which is a new analytical method that combines the Laplace transform with the residual power series method; it was first introduced in [26], and it is implemented by researchers to solve several models of fractional ordinary and partial differential equations and systems. This method shows its efficiency and applicability in solving similar problems [27–35].

The main aim of this article is to present an analytical series solution of the fractional SIR model. We use the LRPSM to get the solution in the form of a rapidly convergent series. We introduce the method and state the convergence analysis of the method, then we apply it to solve the proposed model. Numerical simulations of the results are discussed, and comparisons are made with the results obtained from other numerical methods. The strength of the LRPSM arises in the ability of solving similar models and presenting many terms of the series solutions with fewer calculations and without the need of linearization, discretization, or differentiation as with other numerical methods.

The layout of this article is as follows: in Section 2, we present some definitions and theorems related to fractional power series and the analysis of LRPSM. In Section 3, we construct the series solution of the fractional SIR model in the sense of Caputo's derivative by LRPSM. In Section 4, we introduce some numerical simulations of our results and comparisons to other numerical methods. Finally, the conclusion section is presented in Section 5.

## 2. Fractional Power Series

In this section of the article, we introduce some basic definitions and characteristics of the Caputo fractional derivative and the Laplace transform. Also, we present theorems about the fractional Taylor's series of expansions.

*Definition 1.* Caputo fractional derivative of the function $\varphi(\tau)$ of order $\gamma$ is given by the following equation:

$$D^{\gamma}\varphi(\tau) = \begin{cases} J_{\tau}^{m-\gamma}\varphi^{(r)}(\tau), & r-1<\gamma<r,\ \tau\geq 0,\\ \varphi^{(r)}(\tau), & \gamma = r,\ \tau\geq 0,\end{cases} \tag{3}$$

where $r \in \mathbb{N}$ and $J_{\tau}^{\gamma}$ is the Riemann–Liouville integral of the fractional order $\gamma$ to the function $\varphi(\tau)$, provided the integral exists.

There are many properties of Caputo's derivative, that might be found in [23, 24], and we mention some of them as follows:

(i) $D^{\gamma}c = 0,\ c \in \mathbb{R}$
(ii) $D^{\gamma}t^{\beta} = (\Gamma(\beta+1)/\Gamma(\beta+1-\gamma))\tau^{\beta-\gamma}, \tau\geq 0, \beta> -1, \gamma>0$
(iii) $D^{\gamma}J_{\tau}^{\gamma}\varphi(\tau) = \varphi(\tau)$
(iv) $J_{\tau}^{\gamma}D^{\gamma}\varphi(\tau) = \varphi(\tau) - \sum_{i=0}^{r-1}\varphi^{(i)}(0^+)(\tau^i/i!),\ r-1<\gamma\leq r$

*Definition 2.* Let $\varphi(\tau)$ be a piecewise continuous function on $[0,\infty)$. If $\varphi(\tau)$ is of exponential order, then $\varphi(\tau)$ has the Laplace transform which is defined as follows:

$$\begin{aligned}\Phi(s) &= \mathscr{L}[\varphi(\tau)]\\ &= \int_0^{\infty} e^{-st}\varphi(\tau)\mathrm{d}\tau, \quad s>0.\end{aligned} \tag{4}$$

The inverse Laplace transform is given by

$$\begin{aligned}\varphi(\tau) &= \mathscr{L}^{-1}[\Phi(s)]\\ &= \int_{\alpha-i\infty}^{\alpha+i\infty} e^{st}\Phi(s)\mathrm{d}s,\\ \alpha &= Re(s) > c,\end{aligned} \tag{5}$$

for some, $c$ provided the integral exists.

In the following arguments, we list some of the most popular properties of the Laplace transform.

If $\Phi(s) = \mathscr{L}[\varphi(\tau)]$, $\Psi(s) = \mathscr{L}[\psi(\tau)]$, and $c, d$ are constants, then

(i) $\mathscr{L}[c\varphi(\tau) + d\psi(\tau)] = c\Phi(s) + d\Psi(s)$
(ii) $\mathscr{L}^{-1}[c\Phi(s) + d\Psi(s)] = c\varphi(\tau) + d\psi(\tau)$
(iii) $\lim_{s\longrightarrow\infty} s\Phi(s) = \varphi(0)$
(iv) $\mathscr{L}[D^{\gamma}\varphi(\tau)] = s^{\gamma}\Phi(s) - \sum_{k=0}^{r-1}s^{\gamma-k-1}\varphi^{(k)}(0),\ r-1<\alpha\leq r$
(v) $\mathscr{L}[D^{n\gamma}\varphi(\tau)] = s^{n\gamma}\Phi(s) - \sum_{k=0}^{n-1}s^{(n-k)\gamma-1}D^{k\gamma}\varphi(0),\ 0<\alpha<1$

For the proof, readers can see [2, 4, 23].

The following theorems illustrate some results about Taylor's fractional series and the convergence analysis of the new approach.

**Theorem 1** (see [23]). *Suppose the function $\varphi(\tau)$ has a power series representation at $\tau = \delta$ of the shape:*

$$\varphi(\tau) = \sum_{n=0}^{\infty} \alpha_n (\tau - \delta)^{n\gamma}, \quad 0 < r - 1 < \gamma \leq r,\ 0 < \tau < \xi. \tag{6}$$

If $\varphi(\tau) \in C[\delta, \delta + \xi]$ and $D^{n\gamma}\varphi(\tau) \in C[\delta, \delta + \xi]$ for $n = 0, 1, \ldots$, then we have the following formula for the coefficient $\alpha_n$ in equation (6), that takes the expression as follows:

$$\alpha_n = \frac{D^{n\gamma}\varphi(\delta)}{\Gamma(n\gamma + 1)}, \quad n = 0, 1, 2, \ldots, \subset, \tag{7}$$

where $D^{n\gamma} = \underbrace{D^{\gamma}D^{\gamma}\cdots D^{\gamma}}_{n\text{-times}}$.

Note that putting $\gamma = 1$ in expression (6), we get the usual Taylor power series.

The next theorem illustrates a new form of Theorem 1 in the Laplace space considering $\delta = 0$.

**Theorem 2.** *The fractional power series in (6) has three possibilities for convergence:*

*(i) If $\tau = \delta$, then the series is convergent and the radius of convergent is zero*

*(ii) If $\tau \geq \delta$, the series is convergent and the radius of convergent is $\infty$*

*(iii) If $\delta \leq \tau < \delta + \zeta$, for some possible real number $\zeta$, then the series converges, and if $\tau > \delta + \zeta$, then the series (1) diverges*

**Theorem 3** (see [23]). *If the fractional power series expansion of the function $\Phi(s) = \mathscr{L}[\varphi(\tau)]$ is expressed as follows:*

$$\Phi(s) = \sum_{n=0}^{\infty} \frac{\alpha_n}{s^{n\gamma+1}}, \quad s > 0,\ 0 < \gamma \leq 1, \tag{8}$$

*then the coefficient $\alpha_n$ can be obtained from the following formula:*

$$\alpha_n = D^{n\gamma}\varphi(0). \tag{9}$$

Moreover, the inverse Laplace transform of the series expansion (8) in Theorem 3 has the form:

$$\varphi(\tau) = \sum_{n=1}^{\infty} \frac{D^{n\gamma}\varphi(0)}{\Gamma(n\gamma + 1)}\tau^{n\gamma}, \quad \tau \geq 0,\ 0 < \gamma \leq 1. \tag{10}$$

**Theorem 4** (see [23]) (Convergence analysis). *Let $\Phi(s) = \mathscr{L}[\varphi(\tau)]$ be a function that has the fractional power series in equation (2). If $s\mathscr{L}[D^{(n+1)\gamma}\varphi(\tau)] \leq V$, on $0 < s \leq b$, where $0 < \gamma \leq 1$ and $V > 0$, then the remainder $R_n(s)$ of the series representation (2) has the following bound:*

$$|R_n(s)| \leq \frac{V}{s^{(n+1)\gamma+1}}, \quad 0 < s \leq b. \tag{11}$$

## 3. Construction of Solutions to Fractional SIR Model

*3.1. Fractional SIR Epidemic Model.* There are several phenomena in the real world in engineering and physics, which can be reformulated by fractional initial value problems. Not all of these problems can be solved exactly, so, they are challenging researchers around the whole world. Our aim in this section is to introduce the main idea of the LRPSM in solving systems of nonlinear fractional differential equations that might be difficult to solve by usual techniques.

Now, we present the fractional SIR model:

$$\begin{aligned} D^{\gamma_1}\varphi(\tau) &= -q_1\varphi(\tau)\psi(\tau) + q_3\psi(\tau), \\ D^{\gamma_2}\psi(\tau) &= q_1\varphi(\tau)\psi(\tau) - q_3\psi(\tau) - q_2\psi(\tau), \\ D^{\gamma_3}\phi(\tau) &= q_2\psi(\tau), \end{aligned} \tag{12}$$

where $\gamma_i \in (0, 1]$, $\forall i = 1, 2, 3$, $D^{\gamma_i}$ denotes the Caputo derivative, $q_1, q_2$, and $q_3$ are real positive numbers, $q_1$ denotes the infection rate, $q_2$ denotes the removal rate, and $q_3$ is the recovery rate.

The given initial conditions for system (12) are as follows:

$$\begin{aligned} \varphi(0) &= \alpha, \\ \psi(0) &= \beta, \\ \phi(0) &= \lambda. \end{aligned} \tag{13}$$

Also, we have:

$$\varphi(\tau) + \psi(\tau) + \phi(\tau) = N, \tag{14}$$

and the relation:

$$D^{\gamma_1}\varphi(\tau) + D^{\gamma_2}\psi(\tau) + D^{\gamma_3}\phi(\tau) = 0. \tag{15}$$

The relation (15) gives an extra condition that enable us to solve only two equations of three variables.

*3.2. LRPSM for Solving Fractional SIR Model.* The basic idea of LRPSM is to apply the Laplace transform on the target equations, then define the so-called Laplace residual functions. After that, multiply each equation by $s^{k\alpha+1}$ by the truncated Laplace residual functions and take the limit at infinity to get the required values of the series coefficients recursively.

To get the series solution of the system (12) using the proposed method, we first operate the Laplace transform on both sides of each equation in system (5), to get:

$$\begin{aligned} \mathscr{L}[D^{\gamma_1}\varphi(\tau)] &= -q_1\mathscr{L}[\varphi(\tau)\psi(\tau)] + q_3\mathscr{L}[\psi(\tau)], \\ \mathscr{L}[D^{\gamma_2}\psi(\tau)] &= q_1\mathscr{L}[\varphi(\tau)\psi(\tau)] - q_3\mathscr{L}[\psi(\tau)] - q_2\mathscr{L}[\psi(\tau)], \\ \mathscr{L}[D^{\gamma_3}\phi(\tau)] &= q_2\mathscr{L}[\psi(\tau)]. \end{aligned} \tag{16}$$

Running Laplace transform on system (16), we get:

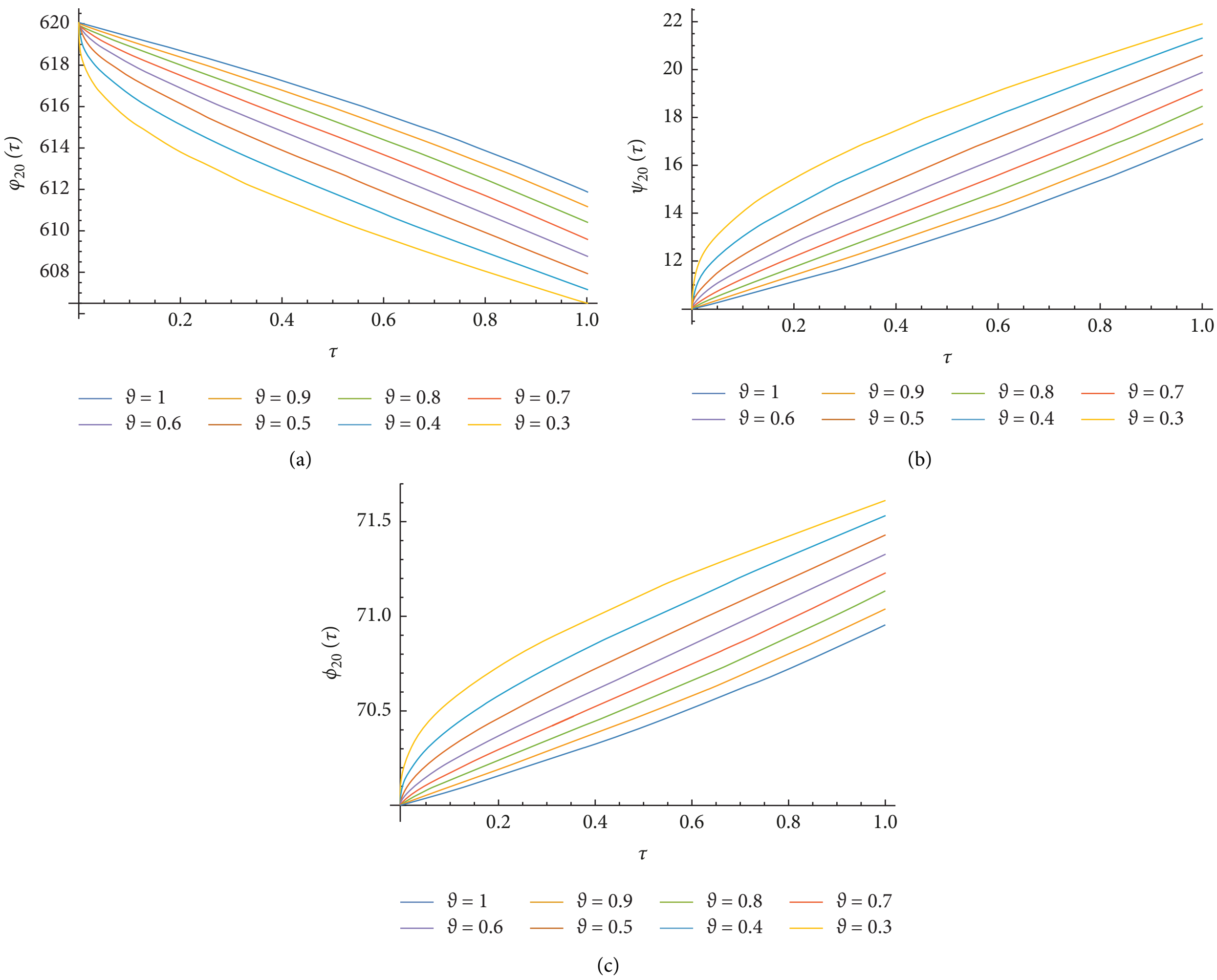

Figure 1: The values of (a) the solution of $\varphi(\tau)$, (b) the solution of $\psi(\tau)$, and (c) the solution of $\phi(\tau)$, for distinct values of $\vartheta\,(\vartheta = 1, 0.9, 0.8, 0.7, 0.6, 0.5, 0.4, 0.3)$ using the LRPSM.

$$
\begin{aligned}
s^{\gamma_1}\Phi(s) - s^{\gamma_1 - 1}\varphi(0) &= -q_1\mathscr{L}\left[\mathscr{L}^{-1}[\Phi(s)]\mathscr{L}^{-1}[\Psi(s)]\right] + q_3\Psi(s),\\
s^{\gamma_2}\Psi(s) - s^{\gamma_1 - 1}\psi(0) &= q_1\mathscr{L}\left[\mathscr{L}^{-1}[\Phi(s)]\mathscr{L}^{-1}[\Psi(s)]\right] - q_3\Psi(s) - q_2\Psi(s),\\
s^{\gamma_3}\mathscr{G}(s) - s^{\gamma_1 - 1}\phi(0) &= q_2\Psi(s).
\end{aligned}
\tag{17}
$$

Simplifying the equations in system (17) and substituting the initial conditions (13), we get:

$$
\begin{aligned}
\Phi(s) &= \frac{\alpha}{s} - \frac{q_1}{s^{\gamma_1}}\mathscr{L}\left[\mathscr{L}^{-1}[\Phi(s)]\mathscr{L}^{-1}[\Psi(s)]\right] + \frac{q_3}{s^{\gamma_1}}\Psi(s),\\
\Psi(s) &= \frac{\beta}{s} + \frac{q_1}{s^{\gamma_2}}\mathscr{L}\left[\mathscr{L}^{-1}[\Phi(s)]\mathscr{L}^{-1}[\Psi(s)]\right] - \frac{q_3}{s^{\gamma_2}}\Psi(s) - q_2\Psi(s),\\
\mathscr{G}(s) &= \frac{\lambda}{s} + \frac{q_2}{s^{\gamma_3}}\Psi(s).
\end{aligned}
\tag{18}
$$

Suppose that the solution of system (18) can be presented in the following series forms:

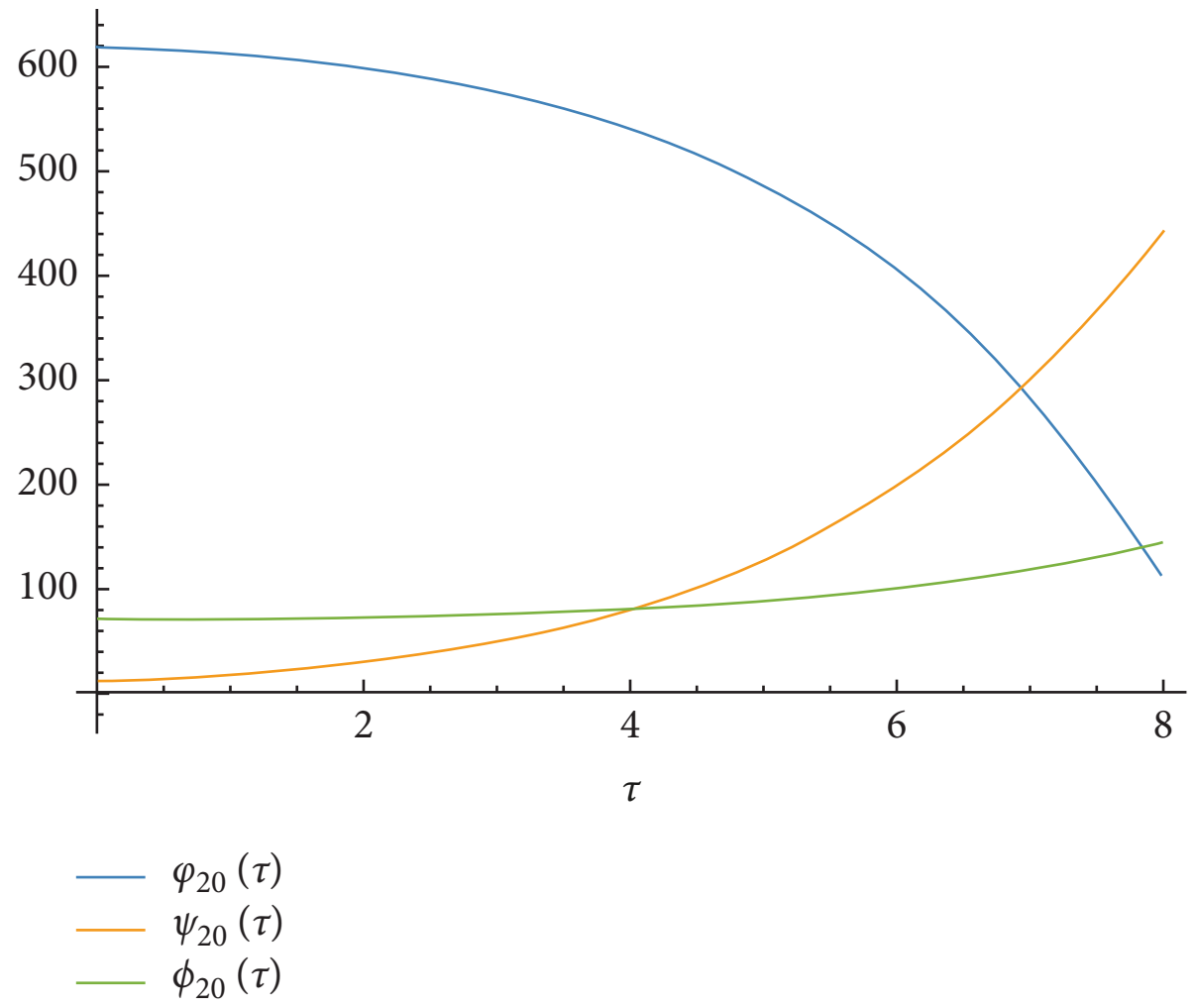

FIGURE 2: The values of $\varphi_{20}(\tau)$, $\psi_{20}(\tau)$, and $\phi_{20}(\tau)$ using LRPSM for $\vartheta = 1$.

$$\Phi(s) = \sum_{n=0}^{\infty} \frac{\alpha_n}{s^{n\gamma_1+1}},$$
$$\Psi(s) = \sum_{n=0}^{\infty} \frac{\beta_n}{s^{n\gamma_2+1}}, \quad (19)$$
$$\mathcal{G}(s) = \sum_{n=0}^{\infty} \frac{\lambda_n}{s^{n\gamma_3+1}}.$$

Using the property that $\lim_{s\longrightarrow\infty} s\,\Phi(s) = \varphi(0)$, it is obvious that

$$\begin{aligned}
\lim_{s\longrightarrow\infty} s\Phi(s) &= \varphi(0) \\
&= \alpha_0 \\
&= \alpha, \\
\lim_{s\longrightarrow\infty} s\Psi(s) &= \psi(0) \\
&= \beta_0 \\
&= \beta, \\
\lim_{s\longrightarrow\infty} s\mathcal{G}(s) &= \phi(0) \\
&= \lambda_0 \\
&= \lambda.
\end{aligned} \quad (20)$$

The series expansion in (19) can be written as follows:

$$\Phi(s) = \frac{\alpha}{s} + \sum_{n=1}^{\infty} \frac{\alpha_n}{s^{n\gamma_1+1}},$$
$$\Psi(s) = \frac{\beta}{s} + \sum_{n=1}^{\infty} \frac{\beta_n}{s^{n\gamma_2+1}}, \quad (21)$$
$$\mathcal{G}(s) = \frac{\lambda}{s} + \sum_{n=1}^{\infty} \frac{\lambda_n}{s^{n\gamma_3+1}}.$$

Now, define the Laplace residual functions of system (18) as follows:

$$\begin{aligned}
\mathcal{L}\mathrm{Res}\Phi(s) &= \Phi(s) - \frac{\alpha}{s} + \frac{q_1}{s^{\gamma_1}} \mathcal{L}\left[\mathcal{L}^{-1}[\Phi(s)]\mathcal{L}^{-1}[\Psi(s)]\right] - \frac{q_3}{s^{\gamma_1}} \Psi(s), \\
\mathcal{L}\mathrm{Res}\Psi(s) &= \Psi(s) - \frac{\beta}{s} - \frac{q_1}{s^{\gamma_2}} \mathcal{L}\left[\mathcal{L}^{-1}[\Phi(s)]\mathcal{L}^{-1}[\Psi(s)]\right] + \frac{q_3}{s^{\gamma_2}} \Psi(s) + q_2 \Psi(s), \\
\mathcal{L}\mathrm{Res}\mathcal{G}(s) &= \mathcal{G}(s) - \frac{\lambda}{s} - \frac{q_2}{s^{\gamma_3}} \Psi(s).
\end{aligned} \quad (22)$$

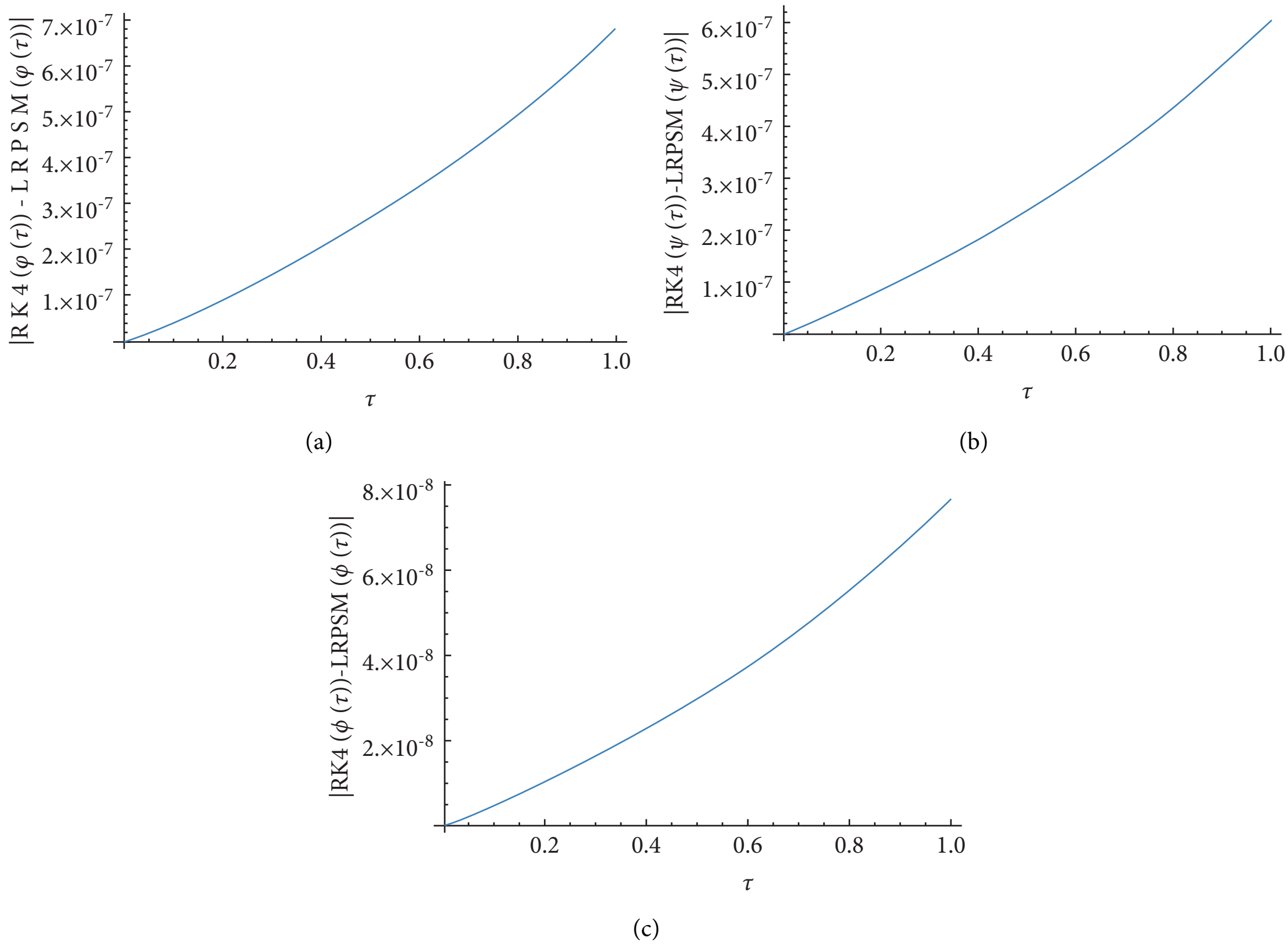

Figure 3: Absolute error between the Runge–Kutta method solution and the LRPSM solutions for $\vartheta = 1$.

The $k$th truncated series of (21) has the form:

$$\begin{aligned}\Phi_k(s) &= \frac{\alpha}{s} + \sum_{n=1}^{k} \frac{\alpha_n}{s^{n\gamma_1+1}},\\ \Psi_k(s) &= \frac{\beta}{s} + \sum_{n=1}^{k} \frac{\beta_n}{s^{n\gamma_2+1}},\\ \mathcal{G}_k(s) &= \frac{\lambda}{s} + \sum_{n=1}^{k} \frac{\lambda_n}{s^{n\gamma_3+1}}.\end{aligned} \tag{23}$$

and the $k$th Laplace residual functions as follows:

$$\begin{aligned}\mathscr{L}\text{Res}_k\Phi(s) &= \Phi_k(s) - \frac{\alpha}{s} + \frac{q_1}{s^{\gamma_1}}\mathscr{L}\left[\mathscr{L}^{-1}[\Phi_k(s)]\mathscr{L}^{-1}[\Psi_k(s)]\right] - \frac{q_3}{s^{\gamma_1}}\Psi_k(s),\\ \mathscr{L}\text{Res}_k\Psi(s) &= \Psi_k(s) - \frac{\beta}{s} - \frac{q_1}{s^{\gamma_2}}\mathscr{L}\left[\mathscr{L}^{-1}[\Phi_k(s)]\mathscr{L}^{-1}[\Psi_k(s)]\right] + \frac{q_3}{s^{\gamma_2}}\Psi_k(s) + q_2\Psi_k(s),\\ \mathscr{L}\text{Res}_k\mathcal{G}(s) &= \mathcal{G}_k(s) - \frac{\lambda}{s} - \frac{q_2}{s^{\gamma_3}}\Psi_k(s).\end{aligned} \tag{24}$$

Now, the first truncated series of (23) are as follows:

Table 1: The value of $\varphi_{20}(\tau)$ for distinct values of $\vartheta(\vartheta = 1, 0.9, 0.8, 0.7, 0.6, 0.5, 0.4, 0.3)$ using the LRPSM.

| $\tau$ | $\vartheta=1$ | $\vartheta=0.9$ | $\vartheta=0.8$ | $\vartheta=0.7$ | $\vartheta=0.6$ | $\vartheta=0.5$ | $\vartheta=0.4$ | $\vartheta=0.3$ |
|---|---|---|---|---|---|---|---|---|
| 0 | 620 | 620 | 620 | 620 | 620 | 620 | 620 | 620 |
| 0.1 | 619.368 | 619.163 | 618.894 | 618.538 | 618.066 | 617.433 | 616.568 | 615.36 |
| 0.2 | 618.702 | 618.384 | 617.991 | 617.501 | 616.886 | 616.107 | 615.109 | 613.811 |
| 0.3 | 618 | 617.596 | 617.112 | 616.529 | 615.823 | 614.961 | 613.901 | 612.594 |
| 0.4 | 617.259 | 616.785 | 616.23 | 615.577 | 614.807 | 613.895 | 612.812 | 611.538 |
| 0.5 | 616.478 | 615.945 | 615.332 | 614.626 | 613.811 | 612.871 | 611.793 | 610.579 |
| 0.6 | 615.655 | 615.071 | 614.411 | 613.664 | 612.821 | 611.872 | 610.818 | 609.686 |
| 0.7 | 614.788 | 614.161 | 613.463 | 612.687 | 611.827 | 610.884 | 609.873 | 608.84 |
| 0.8 | 613.874 | 613.211 | 612.483 | 611.688 | 610.824 | 609.902 | 608.949 | 608.032 |
| 0.9 | 612.911 | 612.218 | 611.469 | 610.663 | 609.808 | 608.919 | 608.04 | 607.251 |
| 1 | 611.897 | 611.181 | 610.418 | 609.612 | 608.774 | 607.933 | 607.14 | 606.494 |

$$\begin{aligned}\Phi_1(s) &= \frac{\alpha}{s} + \frac{\alpha_1}{s^{\gamma_1+1}},\\ \Psi_1(s) &= \frac{\beta}{s} + \frac{\beta_1}{s^{\gamma_2+1}},\\ \mathscr{G}_1(s) &= \frac{\lambda}{s} + \frac{\lambda_1}{s^{\gamma_3+1}}.\end{aligned} \tag{25}$$

Substituting the above values in the first Laplace residual functions to get the following:

$$\begin{aligned}\mathscr{L}\mathrm{Res}_1\Phi(s) &= \frac{\alpha_1}{s^{\gamma_1+1}} + \frac{q_1}{s^{\gamma_1}}\mathscr{L}\left[\mathscr{L}^{-1}\left[\frac{\alpha}{s} + \frac{\alpha_1}{s^{\gamma_1+1}}\right]\mathscr{L}^{-1}\left[\frac{\beta}{s} + \frac{\beta_1}{s^{\gamma_2+1}}\right]\right] - \frac{q_3}{s^{\gamma_1}}\left(\frac{\lambda}{s} + \frac{\lambda_1}{s^{\gamma_3+1}}\right),\\ \mathscr{L}\mathrm{Res}_1\Psi(s) &= \frac{\beta_1}{s^{\gamma_2+1}} - \frac{q_1}{s^{\gamma_2}}\mathscr{L}\left[\mathscr{L}^{-1}\left[\frac{\alpha}{s} + \frac{\alpha_1}{s^{\gamma_1+1}}\right]\mathscr{L}^{-1}\left[\frac{\beta}{s} + \frac{\beta_1}{s^{\gamma_2+1}}\right]\right] + \frac{q_3}{s^{\gamma_2}}\left(\frac{\beta}{s} + \frac{\beta_1}{s^{\gamma_2+1}}\right) + q_2\left(\frac{\beta}{s} + \frac{\beta_1}{s^{\gamma_2+1}}\right),\\ \mathscr{L}\mathrm{Res}_1\mathscr{G}(s) &= \frac{\lambda_1}{s^{\gamma_3+1}} - \frac{q_2}{s^{\gamma_3}}\left(\frac{\beta}{s} + \frac{\beta_1}{s^{\gamma_2+1}}\right).\end{aligned} \tag{26}$$

Using the following facts of the LRPSM, which can be found in [23],

(i) $\mathscr{L}\mathrm{Res}\Phi(s) = 0$

(ii) $\lim_{k\longrightarrow\infty}\mathscr{L}\mathrm{Res}_k\Phi(s) = \mathscr{L}\mathrm{Res}\Phi(s)$, for each $s > 0$

(iii) $\lim_{s\longrightarrow\infty} s\,\mathscr{L}\mathrm{Res}\Phi(s) = 0$ and $\lim_{s\longrightarrow\infty}\mathscr{L}\mathrm{Res}_k\Phi(s) = 0$

(iv) $\lim_{s\longrightarrow\infty}\mathscr{L}\mathrm{Res}_k\Phi(s) = 0$, $k = 1, 2, \ldots$ and $0 < \alpha \leq 1$

Now, multiplying each equation in (26) by $s^{\gamma_1+1}$, $s^{\gamma_2+1}$, and $s^{\gamma_3+1}$, respectively, and then taking the limit as $s \longrightarrow \infty$, we get the first coefficients of the series expansion (23) as follows:

$$\begin{aligned}\alpha_1 &= -q_1\alpha_0\beta_0 + q_3\beta_0\\ &= -q_1\alpha\beta + q_3\beta,\\ \beta_1 &= -\alpha_1 - q_2\beta_0\\ &= q_1\alpha\beta - q_3\beta - q_2\beta,\\ \lambda_1 &= q_2\beta_1.\end{aligned} \tag{27}$$

Repeating the previous steps, we can get other coefficients.

In addition, if we take $\gamma_1 = \gamma_2 = \gamma_3 = \vartheta$, one can get the coefficients of the series solution as follows:

Table 2: The value of $\psi_{20}(\tau)$ for distinct values of $\vartheta(\vartheta = 1, 0.9, 0.8, 0.7, 0.6, 0.5, 0.4, 0.3)$ using the LRPSM.

| $\tau$ | $\vartheta=1$ | $\vartheta=0.9$ | $\vartheta=0.8$ | $\vartheta=0.7$ | $\vartheta=0.6$ | $\vartheta=0.5$ | $\vartheta=0.4$ | $\vartheta=0.3$ |
|---|---|---|---|---|---|---|---|---|
| 0 | 10 | 10 | 10 | 10 | 10 | 10 | 10 | 10 |
| 0.1 | 10.5577 | 10.7388 | 10.9768 | 11.2905 | 11.7069 | 12.2658 | 13.0284 | 14.0934 |
| 0.2 | 11.1457 | 11.4261 | 11.7735 | 12.206 | 12.7482 | 13.4348 | 14.3149 | 15.4584 |
| 0.3 | 11.7658 | 12.122 | 12.549 | 13.0631 | 13.6858 | 14.4455 | 15.379 | 16.5297 |
| 0.4 | 12.4194 | 12.8378 | 13.3274 | 13.9026 | 14.5813 | 15.3851 | 16.3381 | 17.4595 |
| 0.5 | 13.1083 | 13.579 | 14.1194 | 14.7415 | 15.459 | 16.2866 | 17.2358 | 18.3036 |
| 0.6 | 13.8343 | 14.3494 | 14.9312 | 15.5888 | 16.3317 | 17.167 | 18.094 | 19.0894 |
| 0.7 | 14.5993 | 15.1521 | 15.7671 | 16.4505 | 17.207 | 18.0363 | 18.9256 | 19.8329 |
| 0.8 | 15.4052 | 15.9896 | 16.6304 | 17.3307 | 18.0902 | 18.901 | 19.7385 | 20.5441 |
| 0.9 | 16.2541 | 16.8642 | 17.5238 | 18.2326 | 18.9851 | 19.7656 | 20.5383 | 21.2299 |
| 1 | 17.1481 | 17.7782 | 18.4498 | 19.1588 | 19.8945 | 20.6334 | 21.329 | 21.8954 |

$$
\begin{aligned}
\alpha_1 &= \beta(q_3 - \alpha q_1),\\
\alpha_2 &= \beta(\alpha q_1 - q_3)(q_1(\beta-\alpha) + q_2 + q_3),\\
\alpha_3 &= \frac{\beta(\alpha q_1 - q_3)\left(q_1^2\left((\alpha-\beta)^2\Gamma^2(\vartheta+1) - \alpha\beta\,\Gamma(2\vartheta+1)\right)\right)}{\Gamma^2(\vartheta+1)}\\
&+ \frac{\beta(\alpha q_1 - q_3)\left(q_1\left(q_2\left((\beta-2\alpha)\Gamma^2(\vartheta+1) + \beta\Gamma(2\vartheta+1)\right) + q_3\left(\beta\Gamma(2\vartheta+1) - 2(\alpha-\beta)\Gamma^2(\vartheta+1)\right)\right)\right)}{\Gamma^2(\vartheta+1)}\\
&+ \frac{\beta(\alpha q_1 - q_3)(q_2+q_3)^2\Gamma^2(\vartheta+1)}{\Gamma^2(\vartheta+1)}, \ldots,\\
\beta_1 &= \beta(\alpha q_1 - q_2 - q_3),\\
\beta_2 &= \beta\left(\alpha q_1^2(\alpha-\beta) + q_1(q_3(\beta-2\alpha) - 2\alpha q_2) + (q_2+q_3)^2\right),\\
\beta_3 &= \frac{q_1^2\beta\left(q_3\left(2\alpha\beta\Gamma(2\vartheta+1) - \left(3\alpha^2 - 4\alpha\beta + \beta^2\right)\Gamma^2(\vartheta+1)\right) + \alpha q_2\left((2\beta-3\alpha)\Gamma^2(\vartheta+1) + \beta\Gamma(2\vartheta+1)\right)\right)}{\Gamma^2(\vartheta+1)}\\
&+ \frac{\alpha\beta q_1^3\left((\alpha-\beta)^2\Gamma^2(\vartheta+1) - \alpha\beta\Gamma(2\vartheta+1)\right)}{\Gamma^2(\vartheta+1)}\\
&+ \frac{(q_2+q_3)q_1\left(q_3\left((3\alpha-2\beta)\Gamma^2(\vartheta+1) - \beta\Gamma(2\vartheta+1)\right)3\alpha\beta q_2\Gamma^2(\vartheta+1)\right)}{\Gamma^2(\vartheta+1)}\\
&- \frac{(q_2+q_3)^3\Gamma^2(\vartheta+1)}{\Gamma^2(\vartheta+1)}, \ldots,\\
\lambda_1 &= \beta q_2,\\
\lambda_2 &= -\beta q_2(-\alpha q_1 + q_2 + q_3),\\
\lambda_3 &= \beta q_2\left(\alpha q_1^2(\alpha-\beta) + q_1(q_3(\beta-2\alpha) - 2\alpha q_2) + (q_2+q_3)^2\right), \ldots.
\end{aligned}
\tag{28}
$$

The $k$ th coefficients of the series solutions (21) have the form:

Table 3: The value of $\phi_{20}(\tau)$ for distinct values of $\vartheta(\vartheta = 1, 0.9, 0.8, 0.7, 0.6, 0.5, 0.4, 0.3)$ using the LRPSM.

| $\tau$ | $\vartheta=1$ | $\vartheta=0.9$ | $\vartheta=0.8$ | $\vartheta=0.7$ | $\vartheta=0.6$ | $\vartheta=0.5$ | $\vartheta=0.4$ | $\vartheta=0.3$ |
|---|---|---|---|---|---|---|---|---|
| 0 | 70 | 70 | 70 | 70 | 70 | 70 | 70 | 70 |
| 0.1 | 70.074 | 70.098 | 70.1297 | 70.1714 | 70.2269 | 70.3015 | 70.4037 | 70.5469 |
| 0.2 | 70.1521 | 70.1894 | 70.2357 | 70.2934 | 70.3659 | 70.458 | 70.5764 | 70.7311 |
| 0.3 | 70.2346 | 70.2821 | 70.3391 | 70.4079 | 70.4914 | 70.5937 | 70.7199 | 70.8762 |
| 0.4 | 70.3216 | 70.3775 | 70.4431 | 70.5203 | 70.6116 | 70.7203 | 70.8496 | 71.0026 |
| 0.5 | 70.4135 | 70.4765 | 70.5491 | 70.6328 | 70.7297 | 70.842 | 70.9714 | 71.1177 |
| 0.6 | 70.5105 | 70.5796 | 70.6579 | 70.7467 | 70.8475 | 70.9613 | 71.0881 | 71.225 |
| 0.7 | 70.6128 | 70.6872 | 70.7703 | 70.8629 | 70.9659 | 71.0793 | 71.2015 | 71.3268 |
| 0.8 | 70.7208 | 70.7997 | 70.8865 | 70.9818 | 71.0856 | 71.197 | 71.3126 | 71.4244 |
| 0.9 | 70.8347 | 70.9174 | 71.0071 | 71.1039 | 71.2072 | 71.315 | 71.4222 | 71.5187 |
| 1 | 70.955 | 71.0406 | 71.1323 | 71.2296 | 71.3311 | 71.4336 | 71.5307 | 71.6103 |

$$\begin{aligned}
\alpha_{k+1} &= -q_1 \sum_{i=0}^{k} \frac{\alpha_i \beta_{k-i} \Gamma(1+n\vartheta)}{\Gamma(1+i\vartheta)\Gamma(1+(n-i)\vartheta)} + q_3\beta_k,\\
\beta_{k+1} &= -\alpha_{k+1} - q_2\beta_k,\\
\lambda_{k+1} &= q_2\beta_k,\\
&\quad k = 0, 1, 2, \ldots.
\end{aligned} \tag{29}$$

Thus, the $k$th solution of systems (12) and (13) in Laplace space can be expressed as follows:

$$\begin{aligned}
\Phi_k(s) &= \frac{\alpha}{s} + \frac{\beta(q_3 - \alpha q_1)}{s^{\vartheta+1}} + \frac{\beta(\alpha q_1 - q_3)(q_1(\beta-\alpha) + q_2 + q_3)}{s^{2\vartheta+1}} + \cdots\\
&\quad + \frac{1}{s^{k\vartheta+1}}\left(-q_1 \sum_{i=0}^{k} \frac{\alpha_i \beta_{n-i} \Gamma(1+n\vartheta)}{\Gamma(1+i\vartheta)\Gamma(1+(n-i)\vartheta)} + q_3\beta_k\right),\\
\Psi_k(s) &= \frac{\beta}{s} + \frac{\beta(\alpha q_1 - q_2 - q_3)}{s^{\vartheta+1}}\\
&\quad + \frac{\beta\left(\alpha q_1^2(\alpha-\beta) + q_1(q_3(\beta - 2\alpha) - 2\alpha q_2) + (q_2+q_3)^2\right)}{s^{2\vartheta+1}} + \cdots\\
&\quad + \frac{-\alpha_{k+1} - q_2\beta_k}{s^{k\vartheta+1}},\\
\mathcal{G}_k(s) &= \frac{\lambda}{s} + \frac{q_2\beta_1}{s^{\vartheta+1}} + \frac{q_2\beta_2}{s^{2\vartheta+1}} + \cdots + \frac{q_2\beta_k}{s^{k\vartheta+1}}.
\end{aligned} \tag{30}$$

Operating the inverse Laplace transform on each equation in (32), we obtain the $k$th solution of system (12) as follows:

Table 4: The value of $\varphi_{20}(\tau)$ using RK4 and LRPSM for $\vartheta = 1$.

| $\tau_i$ | RK4 | LRPSM | $\lvert RK4(\varphi(\tau)) - LRPSM(\varphi(\tau))\rvert$ |
|---|---|---|---|
| 0 | 620.000000000000000 | 620.000000000000000 | 0 |
| 0.1 | 619.368327701129830 | 619.368327657510350 | $4.36195 \times 10^{-8}$ |
| 0.2 | 618.702153816378770 | 618.702153724652930 | $9.17258 \times 10^{-8}$ |
| 0.3 | 617.999682039910570 | 617.999681895280220 | $1.4463 \times 10^{-7}$ |
| 0.4 | 617.259032603941480 | 617.259032401282640 | $2.02659 \times 10^{-7}$ |
| 0.5 | 616.478239561945540 | 616.478239295794650 | $2.66151 \times 10^{-7}$ |
| 0.6 | 615.655248125904110 | 615.655247790443130 | $3.35461 \times 10^{-7}$ |
| 0.7 | 614.787912080138200 | 614.787911669181200 | $4.10957 \times 10^{-7}$ |
| 0.8 | 613.873991297021010 | 613.873990803999960 | $4.93021 \times 10^{-7}$ |
| 0.9 | 612.911149382834760 | 612.911148800787370 | $5.82047 \times 10^{-7}$ |
| 1 | 611.896951485213090 | 611.896950806769950 | $6.78443 \times 10^{-7}$ |

$$
\begin{aligned}
\varphi_k(\tau) &= \alpha + \frac{\beta(q_3 - \alpha q_1)}{\Gamma(\vartheta+1)}\tau^{\vartheta} + \frac{\beta(\alpha q_1 - q_3)(q_1(\beta-\alpha) + q_2 + q_3)}{\Gamma(2\vartheta+1)}\tau^{2\vartheta} + \cdots \\
&\quad + \frac{\tau^{k\vartheta}}{\Gamma(k\vartheta+1)}\left(-q_1\sum_{i=0}^{k}\frac{\alpha_i\beta_{n-i}\Gamma(1+n\vartheta)}{\Gamma(1+i\vartheta)\Gamma(1+(n-i)\vartheta)} + q_3\beta_k\right), \\
\psi(\tau) &= \beta + \frac{\beta(\alpha q_1 - q_2 - q_3)}{\Gamma(\vartheta+1)}\tau^{\vartheta} \\
&\quad + \frac{\beta\left(\alpha q_1^2(\alpha-\beta) + q_1(q_3(\beta-2\alpha) - 2\alpha q_2) + (q_2+q_3)^2\right)}{\Gamma(2\vartheta+1)}\tau^{2\vartheta} + \cdots \\
&\quad + \frac{-\alpha_{k+1} - q_2\beta_k}{\Gamma(k\vartheta+1)}\tau^{k\vartheta}, \\
\phi_k(\tau) &= \lambda + \frac{q_2\beta_1}{\Gamma(\vartheta+1)}\tau^{\vartheta} + \frac{q_2\beta_2}{\Gamma(2\vartheta+1)}\tau^{2\vartheta} + \cdots + \frac{q_2\beta_k}{\Gamma(k\vartheta+1)}\tau^{k\vartheta}.
\end{aligned}
\tag{31}
$$

As $k \longrightarrow \infty$, the $k$th truncated solutions converge to the exact solutions in the integer orders:

$$
\begin{aligned}
\varphi_k(\tau) &\longrightarrow \varphi(\tau), \\
\psi_k(\tau) &\longrightarrow \psi(\tau), \\
\phi_k(\tau) &\longrightarrow \phi(\tau).
\end{aligned}
\tag{32}
$$

Hence, we get the required solution.

## 4. Numerical Simulation

This section presents the solution of a numerical example of a SIR epidemic system using the LRPSM, and it introduces numerical simulations and figures to show the effectiveness of the suggested approach. The outcomes demonstrate the dependability and strength of LRPSM in solving such problems.

We consider the epidemic SIR model as follows:

$$
\begin{aligned}
D^{\vartheta}\varphi(\tau) &= -0.001\varphi(\tau)\psi(\tau) + 0.005\psi(\tau), \\
D^{\vartheta}\psi(\tau) &= 0.001\varphi(\tau)\psi(\tau) - 0.005\psi(\tau) - 0.072\psi(\tau), \\
D^{\vartheta}\phi(\tau) &= 0.072\psi(\tau), \quad 0 < \vartheta \leq 1,
\end{aligned}
\tag{33}
$$

subject to the initial conditions:

$$
\begin{aligned}
\varphi(0) &= 620, \\
\psi(0) &= 10, \\
\phi(0) &= 70.
\end{aligned}
\tag{34}
$$

The total number of populations is $N = 700$.

To solve system (33) by the proposed method, we apply the Laplace transform on each equation in the system, and using the initial conditions (36), we get:

Table 5: The value of $\psi_{20}(\tau)$ using RK4 and LRPSM for $\vartheta = 1$.

| $\tau_i$ | RK4 | LRPSM | $\lvert RK4(\psi(\tau)) - LRPSM(\psi(\tau))\rvert$ |
|---|---|---|---|
| 0 | 10.000000000000000 | 10.000000000000000 | 0 |
| 0.1 | 10.557682415377140 | 10.557682454124135 | $3.8747 \times 10^{-8}$ |
| 0.2 | 11.145742665297318 | 11.145742746766953 | $8.14696 \times 10^{-8}$ |
| 0.3 | 11.765752718571312 | 11.765752847013921 | $1.28443 \times 10^{-7}$ |
| 0.4 | 12.419356430322715 | 12.419356610275818 | $1.79953 \times 10^{-7}$ |
| 0.5 | 13.108271733277171 | 13.108271969577906 | $2.36301 \times 10^{-7}$ |
| 0.6 | 13.834292759467655 | 13.834293057265034 | $2.97797 \times 10^{-7}$ |
| 0.7 | 14.599291870396474 | 14.599292235163372 | $3.64767 \times 10^{-7}$ |
| 0.8 | 14.599291870396474 | 15.405222008647254 | $4.37545 \times 10^{-7}$ |
| 0.9 | 16.254116280801693 | 16.254116797279011 | $5.16477 \times 10^{-7}$ |
| 1 | 17.148093929799238 | 17.148094531720652 | $6.01921 \times 10^{-7}$ |

$$\begin{aligned}
\Phi(s) &= \frac{620}{s} - \frac{0.001}{s^{\vartheta}}\mathscr{L}\left[\mathscr{L}^{-1}[\Phi(s)]\mathscr{L}^{-1}[\Psi(s)]\right] + \frac{0.005}{s^{\vartheta}}\Psi(s),\\
\Psi(s) &= \frac{10}{s} + \frac{0.001}{s^{\vartheta}}\mathscr{L}\left[\mathscr{L}^{-1}[\Phi(s)]\mathscr{L}^{-1}[\Psi(s)]\right] - \frac{0.005}{s^{\vartheta}}\Psi(s) - \frac{0.072}{s^{\vartheta}}\Psi(s),\\
\mathscr{G}(s) &= \frac{70}{s} + \frac{0.072}{s^{\vartheta}}\Psi(s).
\end{aligned} \tag{35}$$

Now, the $k$th truncated expansion of the solution of system (35) is as follows:

$$\begin{aligned}
\Phi_k(s) &= \frac{620}{s} + \sum_{n=1}^{k}\frac{\alpha_n}{s^{n\vartheta+1}},\\
\Psi_k(s) &= \frac{10}{s} + \sum_{n=1}^{k}\frac{\beta_n}{s^{n\vartheta+1}},\\
\mathscr{G}_k(s) &= \frac{70}{s} + \sum_{n=1}^{k}\frac{\lambda_n}{s^{n\vartheta+1}}.
\end{aligned} \tag{36}$$

We define the $k$th Laplace residual functions as follows:

$$\begin{aligned}
\mathscr{L}\mathrm{Res}_k\Phi(s) &= \Phi_k(s) - \frac{620}{s} + \frac{0.001}{s^{\vartheta}}\mathscr{L}\left[\mathscr{L}^{-1}[\Phi(s)]\mathscr{L}^{-1}[\Psi(s)]\right] - \frac{0.005}{s^{\vartheta}}\Psi_k(s),\\
\mathscr{L}\mathrm{Res}_k\Psi(s) &= \Psi_k(s) - \frac{10}{s} - \frac{0.001}{s^{\vartheta}}\mathscr{L}\left[\mathscr{L}^{-1}[\Phi(s)]\mathscr{L}^{-1}[\Psi(s)]\right] + \frac{0.005}{s^{\vartheta}}\Psi_k(s) + \frac{0.072}{s^{\vartheta}}\Psi_k(s),\\
\mathscr{L}\mathrm{Res}_k\mathscr{G}(s) &= \mathscr{G}_k(s) - \frac{70}{s} - \frac{0.072}{s^{\vartheta}}\Psi_k(s).
\end{aligned} \tag{37}$$

Table 6: The value of $\phi_{20}(\tau)$ using RK4 and LRPSM for $\vartheta = 1$.

| $\tau_i$ | RK4 | LRPSM | $\lvert \mathrm{RK4}(\phi(\tau)) - \mathrm{LRPSM}(\phi(\tau)) \rvert$ |
|---|---|---|---|
| 0 | 70.000000000000000 | 70.000000000000000 | 0 |
| 0.1 | 70.073989883493013 | 70.073989888365404 | $4.87239 \times 10^{-9}$ |
| 0.2 | 70.152103518323841 | 70.152103528579929 | $1.02561 \times 10^{-8}$ |
| 0.3 | 70.234565241518069 | 70.234565257705697 | $1.61876 \times 10^{-8}$ |
| 0.4 | 70.321610965735758 | 70.321610988441307 | $2.27055 \times 10^{-8}$ |
| 0.5 | 70.413488704777279 | 70.413488734627379 | $2.98501 \times 10^{-8}$ |
| 0.6 | 70.510459114628276 | 70.510459152291887 | $3.76636 \times 10^{-8}$ |
| 0.7 | 70.612796049465331 | 70.612796095655639 | $4.61903 \times 10^{-8}$ |
| 0.8 | 70.720787131876406 | 70.720787187352826 | $5.54764 \times 10^{-8}$ |
| 0.9 | 70.834734336363510 | 70.834734401933645 | $6.55701 \times 10^{-8}$ |
| 1 | 70.954954584987661 | 70.954954661509220 | $7.65216 \times 10^{-8}$ |

Multiplying each equation in (39) by $s^{k\vartheta+1}, k = 1, 2, \cdots$ recursively, and taking the limit as $s \longrightarrow \infty$, we get the coefficients of the series solutions as follows:

$$
\begin{aligned}
\alpha_1 &= -6.15,\\
\beta_1 &= 5.43,\\
\lambda_1 &= 0.72,\\
\alpha_2 &= -3.27795,\\
\beta_2 &= 2.88699,\\
\lambda_2 &= 0.39096,\\
\alpha_3 &= \frac{0.03339\Gamma(2\vartheta+1)}{\Gamma^2(\vartheta+1)} - 1.74272,\\
\beta_3 &= 1.53486 - \frac{0.0334\Gamma(2\vartheta+1)}{\Gamma^2(\vartheta+1)},\\
\lambda_3 &= 0.20786,\\
\alpha_4 &= -0.92651 + \frac{0.02021\Gamma(2\vartheta+1)}{\Gamma^2(\vartheta+1)} + \frac{0.03555\Gamma(3\vartheta+1)}{\Gamma(\vartheta+1)\Gamma(2\vartheta+1)},\\
\beta_4 &= 0.816 - \frac{0.0178\Gamma(2\vartheta+1)}{\Gamma^2(\vartheta+1)} - \frac{0.03556\Gamma(3\vartheta+1)}{\Gamma(\vartheta+1)\Gamma(2\vartheta+1)},\\
\lambda_4 &= 0.11051 - \frac{0.00241\Gamma(2\vartheta+1)}{\Gamma^2(\vartheta+1)},\\
\alpha_5 &= -0.49257 - \frac{0.00039\Gamma(4\vartheta+1)\Gamma(2\vartheta+1)}{\Gamma^3(\vartheta+1)\Gamma(3\vartheta+1)} + \frac{0.01074\Gamma(2\vartheta+1)}{\Gamma^2(\vartheta+1)}\\
&\quad + \frac{0.02151\Gamma(3\vartheta+1)}{\Gamma(2\vartheta+1)} + \frac{0.0189\Gamma(4\vartheta+1)}{\Gamma(3\vartheta+1)} + \frac{0.00946\Gamma(4\vartheta+1)}{\Gamma^2(2\vartheta+1)},\\
\beta_5 &= 0.43382 + \frac{0.00039\Gamma(4\vartheta+1)\Gamma(2\vartheta+1)}{\Gamma^3(\vartheta+1)\Gamma(3\vartheta+1)} - \frac{0.00946\Gamma(2\vartheta+1)}{\Gamma^2(\vartheta+1)}\\
&\quad - \frac{0.01895\Gamma(3\vartheta+1)}{\Gamma(2\vartheta+1)\Gamma(\vartheta+1)} - \frac{0.018902\Gamma(4\vartheta+1)}{\Gamma(3\vartheta+1)\Gamma(\vartheta+1)} - \frac{0.009463\Gamma(4\vartheta+1)}{\Gamma^2(2\vartheta+1)},\\
\lambda_5 &= 0.058752 - \frac{0.001282\Gamma(2\vartheta+1)}{\Gamma^2(\vartheta+1)} - \frac{0.00256\Gamma(3\vartheta+1)}{\Gamma(\vartheta+1)\Gamma(2\vartheta+1)},
\end{aligned} \tag{38}
$$

$$
\begin{aligned}
\varphi_6(\tau) = {} & 620 - \frac{6.15\tau^{\vartheta}}{\Gamma(1+\vartheta)} - \frac{3.27795\tau^{2\vartheta}}{\Gamma(1+2\vartheta)} + \left(\frac{0.03339\Gamma(2\vartheta+1)}{\Gamma^2(\vartheta+1)} - 1.74272\right)\frac{\tau^{3\vartheta}}{\Gamma(1+3\vartheta)} \\
& + \left(-0.92651 + \frac{0.02021\Gamma(2\vartheta+1)}{\Gamma^2(\vartheta+1)} + \frac{0.03555\Gamma(3\vartheta+1)}{\Gamma(\vartheta+1)\Gamma(2\vartheta+1)}\right)\frac{\tau^{4\vartheta}}{\Gamma(1+4\vartheta)} \\
& + \left(-0.49257 - \frac{0.00039\Gamma(4\vartheta+1)\Gamma(2\vartheta+1)}{\Gamma^3(\vartheta+1)\Gamma(3\vartheta+1)} + \frac{0.01074\Gamma(2\vartheta+1)}{\Gamma^2(\vartheta+1)} + \frac{0.02151\Gamma(3\vartheta+1)}{\Gamma(2\vartheta+1)} + \frac{0.0189\Gamma(4\vartheta+1)}{\Gamma(3\vartheta+1)} \right. \\
& \left. + \frac{0.00946\Gamma(4\vartheta+1)}{\Gamma^2(2\vartheta+1)}\right)\frac{\tau^{5\vartheta}}{\Gamma(1+5\vartheta)} \\
& + \left(\begin{array}{c}
-0.261876 + \dfrac{0.0057256\Gamma(4\vartheta+1)}{\Gamma^2(2\vartheta+1)} + \dfrac{0.010062\Gamma(5\vartheta+1)}{\Gamma(2\vartheta+1)\Gamma(3\vartheta+1)} \\
+ \dfrac{(0.011439\Gamma(3\vartheta+1)/\Gamma(2\vartheta+1)) + (0.010049\Gamma(5\vartheta+1)/\Gamma(4\vartheta+1)) + (0.011436\Gamma(4\vartheta+1)/\Gamma(3\vartheta+1))}{\Gamma(\alpha+1)} \\
+ \dfrac{0.005712\Gamma(2\vartheta+1) - (0.000209\Gamma(5\vartheta+1)/\Gamma(3\vartheta+1)) - (0.000412\Gamma(3\vartheta+1)\Gamma(5\vartheta+1)/\Gamma(4\vartheta+1)\Gamma(2\vartheta+1))}{\Gamma^2(\vartheta+1)} \\
+ \dfrac{\Gamma(2\vartheta+1)\left(-\left(0.000234\Gamma^2(4\vartheta+1)/\Gamma(3\vartheta+1)\right) - 0.000219\Gamma(5\vartheta+1)\right)}{\Gamma^3(\vartheta+1)\Gamma(4\vartheta+1)}
\end{array}\right) \\
& \cdot \frac{\tau^{6\vartheta}}{\Gamma(1+6\vartheta)} + \ldots,
\end{aligned}
\tag{39}
$$

$$
\begin{aligned}
\psi_6(\tau) = {} & 10 + \frac{5.43\tau^{\vartheta}}{\Gamma(1+\vartheta)} + \frac{2.88699\tau^{2\vartheta}}{\Gamma(1+2\vartheta)} + \left(1.53486 - \frac{0.0334\Gamma(2\vartheta+1)}{\Gamma^2(\vartheta+1)}\right)\frac{\tau^{3\vartheta}}{\Gamma(1+3\vartheta)} \\
& + \left(0.816 - \frac{0.0178\Gamma(2\vartheta+1)}{\Gamma^2(\vartheta+1)} - \frac{0.03556\Gamma(3\vartheta+1)}{\Gamma(\vartheta+1)\Gamma(2\vartheta+1)}\right)\frac{\tau^{4\vartheta}}{\Gamma(1+4\vartheta)} \\
& + \left(0.43382 + \frac{0.00039\Gamma(4\vartheta+1)\Gamma(2\vartheta+1)}{\Gamma^3(\vartheta+1)\Gamma(3\vartheta+1)} - \frac{0.00946\Gamma(2\vartheta+1)}{\Gamma^2(\vartheta+1)} - \frac{0.01895\Gamma(3\vartheta+1)}{\Gamma(2\vartheta+1)\Gamma(\vartheta+1)} - \frac{0.018902\Gamma(4\vartheta+1)}{\Gamma(3\vartheta+1)\Gamma(\vartheta+1)} \right. \\
& \left. - \frac{0.009463\Gamma(4\vartheta+1)}{\Gamma^2(2\vartheta+1)}\right)\frac{\tau^{5\vartheta}}{\Gamma(1+5\vartheta)} \\
& + \left(\begin{array}{c}
0.23064 - \dfrac{0.005044\Gamma(4\vartheta+1)}{\Gamma^2(2\vartheta+1)} - \dfrac{0.010062\Gamma(5\vartheta+1)}{\Gamma(2\vartheta+1)\Gamma(3\vartheta+1)} \\
+ \dfrac{-(0.010075\Gamma(3\vartheta+1)/\Gamma(2\vartheta+1)) - (0.010049\Gamma(5\vartheta+1)/\Gamma(4\vartheta+1)) - (0.010075\Gamma(4\vartheta+1)/\Gamma(3\vartheta+1))}{\Gamma(\alpha+1)} \\
+ \dfrac{-0.005031\Gamma(2\vartheta+1) + (0.000206\Gamma(5\vartheta+1)/\Gamma(3\vartheta+1)) + (0.000412\Gamma(3\vartheta+1)\Gamma(5\vartheta+1)/\Gamma(4\vartheta+1)\Gamma(2\vartheta+1))}{\Gamma^2(\vartheta+1)} \\
+ \dfrac{\Gamma(2\vartheta+1)\left(\left(0.000206\Gamma^2(4\vartheta+1)/\Gamma(3\vartheta+1)\right) + 0.00022\Gamma(5\vartheta+1)\right)}{\Gamma^3(\vartheta+1)\Gamma(4\vartheta+1)}
\end{array}\right) \\
& \cdot \frac{\tau^{6\vartheta}}{\Gamma(1+6\vartheta)} + \ldots,
\end{aligned}
$$

$$\phi_6(\tau) = 70 + \frac{0.72\tau^{\vartheta}}{\Gamma(1+\vartheta)} + \frac{0.39096\tau^{2\vartheta}}{\Gamma(1+2\vartheta)} + \frac{0.20786\tau^{3\vartheta}}{\Gamma(1+3\vartheta)} + \left(0.11051 - \frac{0.00241\Gamma(2\vartheta+1)}{\Gamma^2(\vartheta+1)}\right)\frac{\tau^{4\vartheta}}{\Gamma(1+4\vartheta)}$$
$$+\left(0.058752 - \frac{0.001282\Gamma(2\vartheta+1)}{\Gamma^2(\vartheta+1)} - \frac{0.00256\Gamma(3\vartheta+1)}{\Gamma(\vartheta+1)\Gamma(2\vartheta+1)}\right)\frac{\tau^{5\vartheta}}{\Gamma(1+5\vartheta)}$$
$$+\left(\begin{array}{c} 0.031235 + \left(0.000028\Gamma(4\vartheta+1)\Gamma(2\vartheta+1)/\Gamma^3(\vartheta+1)\Gamma(3\vartheta+1)\right) - \left(0.000681\Gamma(2\vartheta+1)/\Gamma^2(\vartheta+1)\right) \\ + \dfrac{-(0.001364\Gamma(3\vartheta+1)/\Gamma(2\vartheta+1)) - (0.001361\Gamma(4\vartheta+1)/\Gamma(3\vartheta+1))}{\Gamma(\vartheta+1)} - \dfrac{0.000681\Gamma(4\vartheta+1)}{\Gamma^2(2\vartheta+1)} \end{array}\right)$$
$$\cdot \frac{\tau^{6\vartheta}}{\Gamma(1+6\vartheta)} + \cdots. \quad (40)$$

The 20th terms solution of systems (33) and (34) at $\vartheta = 1$ is given by the following equation:

$$\begin{aligned}
\varphi_{20}(\tau) &= 620 - 6.15\tau - 1.63898\tau^2 - 0.27932\tau^3 - 0.03248\tau^4 - 0.00231\tau^5 \\
&+ 6.631\times10^{-7}\tau^6 + 2.85675\times10^{-5}\tau^7 + 4.90414\times10^{-6}\tau^8 \\
&+ 4.94538\times10^{-7}\tau^9 + 2.5347\times10^{-8}\tau^{10} - 1.6516\times10^{-9}\tau^{11} \\
&- 5.91732\times10^{-10}\tau^{12} - 8.16453\times10^{-11}\tau^{13} - 6.84287\times10^{-12}\tau^{14} \\
&- 1.83575\times10^{-13}\tau^{15} + 5.1703\times10^{-14}\tau^{16} + 1.11303\times10^{-14}\tau^{17} \\
&+ 1.28567\times10^{-15}\tau^{18} + 8.57039\times10^{-17}\tau^{19} - 9.08862\times10^{-19}\tau^{20}, \\
\psi_{20}(\tau) &= 10 + 5.43\tau + 1.44349\tau^2 + 0.24468\tau^3 + 0.028078\tau^4 + 0.00191\tau^5 \\
&- 2.3538\times10^{-5}\tau^6 - 2.83254\times10^{-5}\tau^7 - 4.64921\times10^{-6}\tau^8 \\
&- 4.57345\times10^{-7}\tau^9 - 2.20542\times10^{-8}\tau^{10} + 1.79596\times10^{-9}\tau^{11} \\
&+ 5.80956\times10^{-10}\tau^{12} + 7.84277\times10^{-11}\tau^{13} + 6.43953\times10^{-12}\tau^{14} \\
&+ 1.52665\times10^{-13}\tau^{15} - 5.239\times10^{-14}\tau^{16} - 1.090842\times10^{-14}\tau^{17} \\
&- 1.24203\times10^{-15}\tau^{18} - 8.09972\times10^{-17}\tau^{19} + 1.20045\times10^{-18}\tau^{20}, \\
\phi_{20}(\tau) &= 70 + 0.72\tau + 0.19548\tau^2 + 0.03464\tau^3 + 0.0044\tau^4 + 0.00041\tau^5 \\
&+ 2.28749\times10^{-5}\tau^6 - 2.42105\times10^{-7}\tau^7 - 2.549283\times10^{-7}\tau^8 \\
&- 3.719366\times10^{-8}\tau^9 - 3.29288\times10^{-9}\tau^{10} - 1.443545\times10^{-10}\tau^{11} \\
&+ 1.07757\times10^{-11}\tau^{12} + 3.2176\times10^{-12}\tau^{13} + 4.0334\times10^{-13}\tau^{14} \\
&+ 3.09097\times10^{-14}\tau^{15} + 6.86992\times10^{-16}\tau^{16} - 2.21887\times10^{-16}\tau^{17} \\
&- 4.36337\times10^{-17}\tau^{18} - 4.70665\times10^{-18}\tau^{19} - 2.9159\times10^{-19}\tau^{20}.
\end{aligned} \quad (41)$$

The following figures illustrate the 20th solution of systems (33) and (34) with various values of $(\vartheta = 1, 0.9, 0.8, 0.7, 0.6, 0.5, 0.4)$ and 0.3.

Figure 1 of $\varphi_{20}(\tau), \psi_{20}(\tau)$, and $\phi_{20}(\tau)$ shows the strength of LRPSM in solving the proposed model, which is obvious, with $\vartheta$ increasing from 0 to one, the solution is stable and coincides with the exact solution obtained in the case $\vartheta = 1$.

In Figure 2 below, we sketch the graph of $\varphi_{20}(\tau), \psi_{20}(\tau)$, and $\phi_{20}(\tau)$.

The following tables, Tables 1–3 present the LRPSM solution of systems (33) and (34) at various values of $\vartheta(\vartheta = 1, 0.9, 0.8, 0.7, 0.6, 0.5, 0.4, 0.3)$. In Table 1, we propose different values of $\varphi_{20}(\tau)$ with different values of $\vartheta$, in Table 2, we propose different values of $\psi_{20}(\tau)$ with different values of $\vartheta$, and in Table 3, we propose different values of $\phi_{20}(\tau)$ with different values of $\vartheta$. We notice the efficiency and strength of the method from the agreement of the values, and they all coincide and converge to the exact solution in the integer order when $\vartheta = 1$.

In the following, we introduce Figure 3, which illustrates the absolute error between the Runge–Kutta method solution and the LRPSM solution at $\vartheta = 1$.

The following tables (Tables 4–6), present comparisons between the obtained results $\varphi_{20}(\tau)$, $\psi_{20}(\tau)$, and $\phi_{20}(\tau)$ from the LRPSM and the fourth-order Runge–Kutta method (RK4). These comparisons prove the strength and convergence of the presented method. We notice from Tables 4–6 the efficiency of the proposed method, since the difference between the methods is too small.

## 5. Conclusions

In this study, we introduce the fractional SIR epidemic model in the sense of Caputo's fractional derivative. The LRPSM is used to solve the proposed system. We present a series of solutions of the model and compare our results with those obtained by the fourth-order Runge–Kutta method. In addition, we sketch the graphs of the solutions with different values of $\vartheta$ and analyze the results. Finally, we conclude that the LRPSM is efficient in solving the SIR epidemic system and similar models. Moreover, as the method is applicable and easy in treating similar problems, it could provide many terms of the series solution with less calculations and effort compared to other numerical methods. In the future, we intend to solve new models by LRPSM and make comparisons to other analytical methods. We conclude the following points from our study:

(i) LRPSM is a powerful technique for solving fractional models

(ii) LRPSM is simple and could provide many terms of the series solution without requiring discretization, linearization, or special assumptions on the conditions

(iii) The method could give exact solutions when the exact one is a polynomial

## Data Availability

No underlying data were collected or produced in this study.

## Conflicts of Interest

The authors declare that they have no conflicts of interest.